\documentclass[10pt,conference]{IEEEtran}
\ifCLASSINFOpdf
\else
\fi

\usepackage{amsthm}
\theoremstyle{definition}
\newtheorem{definition}{Definition}[section] 

\usepackage{xcolor,soul,framed} 
\usepackage{amssymb}
\newcommand{\coloneqq}{\mathrel{\mathop:}=}

\colorlet{shadecolor}{yellow}
\usepackage[pdftex]{graphicx}
\graphicspath{{../pdf/}{../jpeg/}}
\DeclareGraphicsExtensions{.pdf,.jpeg,.png}

\usepackage[cmex10]{amsmath}
\usepackage{mathrsfs}
\usepackage{mdwmath}
\usepackage{mdwtab}
\usepackage{mathtools}
\usepackage{eqparbox}
\usepackage{url}
\usepackage{breakurl}
\usepackage{stmaryrd}
\usepackage{booktabs}
\usepackage{CJK}
\usepackage{algorithm}
\usepackage{algpseudocode}
\usepackage[caption=false, font=normalsize, labelfont=sf, textfont=sf]{subfig}
\usepackage{cite}
\usepackage{makecell}
\usepackage{adjustbox}
\usepackage{hyperref}
\usepackage{algpseudocode}

\begin{document}

%
\title{VeriLoRA: Fine-Tuning Large Language Models with Verifiable Security via Zero-Knowledge Proofs}

\author{\IEEEauthorblockN{Guofu~Liao}
	\IEEEauthorblockA{Shenzhen University\\
		liaoguofu2022@email.szu.edu.cn}
	\and
	\IEEEauthorblockN{Taotao~Wang\textsuperscript{†}}
	\IEEEauthorblockA{Shenzhen University\\
		ttwang@szu.edu.cn}
	\and
	\IEEEauthorblockN{Shengli Zhang\textsuperscript{†}}
	\IEEEauthorblockA{Shenzhen University\\
		zsl@szu.edu.cn}
        \and
        
	\IEEEauthorblockN{Jiqun Zhang}
	\IEEEauthorblockA{Shenzhen University\\
		2453043004@mails.szu.edu.cn}
        \and
        
	\IEEEauthorblockN{Long Shi}
	\IEEEauthorblockA{Nanjing University of Science and Technology\\
		slong1007@gmail.com}
        \and
	\IEEEauthorblockN{Dacheng Tao}
	\IEEEauthorblockA{Nanyang Technological University\\
		dacheng.tao@gmail.com}

\thanks{\textsuperscript{†} Corresponding author.}
\thanks{This paper has been accepted for publication at the Network and Distributed System Security Symposium (NDSS) 2026. }
        }


%


\IEEEoverridecommandlockouts

\maketitle
\pagestyle{plain} 
\begin{abstract}

Fine-tuning large language models (LLMs) is crucial for adapting them to specific tasks, yet it remains computationally demanding and raises concerns about correctness and privacy, particularly in untrusted environments. Although parameter-efficient methods like Low-Rank Adaptation (LoRA) significantly reduce resource requirements, ensuring the security and verifiability of fine-tuning under zero-knowledge constraints remains an unresolved challenge. To address this, we introduce VeriLoRA, the first framework to integrate LoRA fine-tuning with zero-knowledge proofs (ZKPs), achieving provable security and correctness. VeriLoRA employs advanced cryptographic techniques---such as lookup arguments, sumcheck protocols, and polynomial commitments---to verify both arithmetic and non-arithmetic operations in Transformer-based architectures. The framework provides end-to-end verifiability for forward propagation, backward propagation, and parameter updates during LoRA fine-tuning, while safeguarding the privacy of model parameters and training data. Leveraging GPU-based implementations, VeriLoRA demonstrates practicality and efficiency through experimental validation on open-source LLMs like LLaMA, scaling up to 13 billion parameters. By combining parameter-efficient fine-tuning with ZKPs, VeriLoRA bridges a critical gap, enabling secure and trustworthy deployment of LLMs in sensitive or untrusted environments.

\end{abstract}


%
\IEEEpeerreviewmaketitle
\section{Introduction}

The rapid advancement of large language models (LLMs) has transformed natural language processing (NLP), enabling breakthroughs in tasks such as text generation, summarization, and machine translation \cite{ray2023chatgpt,zhang2022optopenpretrainedtransformer,touvron2023llama,11164571,11117175}. However, fine-tuning these  massive models for specific tasks remains a significant challenge due to their enormous parameter sizes and associated computational costs \cite{chowdhery2023palm,touvron2023llama}. To address this, techniques like Low-Rank Adaptation (LoRA) \cite{hu2022lora} have emerged, offering a parameter-efficient way to fine-tune LLMs by introducing trainable low-rank matrices that adapt pretrained weights without modifying the base model. LoRA achieves high performance with significantly reduced resource requirements, making it a practical solution for many applications.

Despite these advancements, the security and correctness of fine-tuning processes remain an open problem, especially in scenarios involving sensitive data or untrusted environments \cite{ghodsi2017safetynets,zhao2021veriml}. For example, when fine-tuning is conducted on proprietary models or outsourced to third-party platforms, there is a critical need to ensure that the computations are performed correctly without exposing the model parameters or training data \cite{11082313}. Existing research has explored zero-knowledge proofs (ZKPs) to verify computations in cryptographic and machine learning contexts, such as inference or simple training tasks. Frameworks like zkML \cite{chen2024zkml} and zkLLM \cite{sun2024zkllm} have demonstrated the feasibility of applying ZKPs to machine learning, but these works primarily focus on inference or small-scale models. The problem of achieving zero-knowledge verifiability for fine-tuning large-scale LLMs, particularly under parameter-efficient methods like LoRA, remains unexplored and presents significant technical challenges.

The task of achieving zero-knowledge verifiability for fine-tuning LLMs is non-trivial due to several reasons:  
\begin{itemize}
    \item \textbf{Computational Complexity}: Fine-tuning involves forward and backward propagation through Transformer-based LLM architectures, which include both arithmetic operations (e.g., matrix multiplications) and non-arithmetic operations (e.g., Softmax normalization, SwiGLU activations). Verifying these computations under zero-knowledge constraints requires novel techniques to handle both types of operations efficiently.
    \item \textbf{Scalability}: Modern LLMs, such as LLaMA-2, can have billions of parameters. Ensuring verifiability without incurring prohibitive computational overhead is a significant challenge, particularly when handling large-scale models and datasets.
    \item \textbf{Non-Arithmetic Operations}: The computations over many critical components of Transformer architectures (such as Softmax and SwiGLU) and the gradient computations over these components involve non-linear and non-arithmetic operations. It is difficult to encode these operations into traditional ZKPs' constrain systems for proving.
    \item \textbf{Privacy and Trust}: While ensuring computational integrity, it is equally important to preserve the privacy of sensitive information, such as the model parameters and training data, which are often proprietary or confidential.
\end{itemize}

To the best of our knowledge, there is no existing work that provides zero-knowledge verifiability for the fine-tuning of large-scale LLMs using LoRA or similar parameter-efficient methods. Addressing this gap is both technically challenging and crucial for enabling secure and trustworthy deployment of LLMs in sensitive or trustless environments. To tackle these challenges, we propose \textbf{VeriLoRA}, a novel framework that integrates LoRA fine-tuning with zero-knowledge proofs to achieve verifiable security and correctness during the fine-tuning of LLMs. The key contributions and novelties of this work are summarized as follows:

\begin{enumerate}
    \item \textbf{First Zero-Knowledge Verifiable Fine-Tuning Framework}: VeriLoRA is the first framework to provide end-to-end zero-knowledge verifiability for the fine-tuning of large-scale LLMs. It ensures that all computational steps in LoRA fine-tuning, including forward propagation, backward propagation, and parameter updates, are provably correct without revealing sensitive information such as model parameters or training data.

    \item \textbf{Innovative Handling of Non-Arithmetic Operations}: VeriLoRA addresses the challenge of verifying non-arithmetic operations occurring in the forward and backward computation process through Transformer layers of LLMs. By leveraging lookup-based arguments, VeriLoRA encodes these operations into zero-knowledge proof systems, making them verifiable under stringent privacy constraints.



    \item \textbf{Experimental Validation on Large-Scale Models}: We demonstrate the practicality of VeriLoRA by applying it to fine-tune open-source LLMs, such as LLaMA-2, using real-world datasets. The experimental results showcase the framework’s ability to achieve verifiable fine-tuning with resaonable computational overhead, paving the way for more secure and trustworthy applications of LLMs.
\end{enumerate}
The structure of the remainder of this paper is as follows.  Section \ref{sec:Preliminary} provides preliminary knowledge of zero knowledge. Section \ref{llm} outlines the typical structure of LLMs. Section \ref{sec:VeriLoRA} details the design of VeriLoRA protocol. Section \ref{sec:Security Analysis} analyzes the security of VeriLoRA protocol. Section \ref{Experimental} presents our experimental setup and results. Section \ref{related work} reviews related work. Section \ref{conclusion} concludes this work and discusses directions for future research.


\section{Preliminary}\label{sec:Preliminary}

This section introduces the foundational background theories and techniques employed by VeriLoRA.

\subsection{Proofs, Arguments and Polynomial Commitments}
\subsubsection{Proofs and Arguments}
In cryptography, proofs and arguments are protocols allowing a prover to convince a verifier of a statement's validity. Proofs provide \emph{information-theoretic soundness}, secure even against unbounded provers, whereas arguments offer \emph{computational soundness}, secure against probabilistic polynomial-time (PPT) adversaries based on cryptographic assumptions. A proof or argument system has three essential properties:
\begin{itemize}
\item \textbf{Completeness}: Honest provers can convince verifiers of true statements with high probability $1 - \mathsf{negl}(\lambda)$.
\item \textbf{Soundness}: Malicious provers cannot convince verifiers of false statements except with negligible probability $\mathsf{negl}(\lambda)$.
\item \textbf{Zero-Knowledge}: Verifiers learn no additional information beyond the truth of the statement (for a formal definition, see Definition~\ref{def:zk}).
\end{itemize}
where $\lambda$ denotes the security parameter, and $\mathsf{negl}(\lambda)$ denotes a negligible function, rapidly approaching zero as $\lambda$ grows.

Interactive proof systems enable a prover to persuade a verifier $\mathcal{V}$ of the truth of a statement $x \in L$, where $L$ is an NP language, via a series of message exchanges. The definition of interactive proof systems is presented as follows:

\begin{definition}[Interactive Proof System]
\label{def:ip}
For an NP language \( L \subseteq \{0,1\}^\ast \) with an associated relation \( R = \{(x,w)\} \), an interactive proof system is a pair of probabilistic polynomial-time (PPT) algorithms \((\mathcal{P}, \mathcal{V})\), which are termed the prover and verifier, respectively.  The protocol of \((\mathcal{P}, \mathcal{V})\) must satisfy the following two requirements. \textbf{Completeness}: If the statement $x$ is true (meaning a valid witness $w$ exists and is known to an honest prover), we have $ \Pr\left[\langle \mathcal{P}(x,w), \mathcal{V}(x) \rangle = 1\right] \geq 1 - \mathsf{negl}(\lambda)$ for every honest prover $\mathcal{P}$ that follows the protocol properly. \textbf{Soundness}: If the statement $x$ is false ($x \notin L$), we have $ \Pr\left[\langle \mathcal{P}^{*}(x,w), \mathcal{V}(x) \rangle = 1\right] \leq \mathsf{negl}(\lambda)$ for any PPT prover $\mathcal{P}^{*}$, even one that deviates maliciously from the protocol.
\end{definition}

We say that an interactive proof is public coin if the verifier’s challenge in each round is independent of prover’s messages in previous rounds. Originally introduced by \cite{goldwasser2019knowledge}, these systems allow resource-constrained verifiers to efficiently validate claims, such as NP-hard computations, while maintaining privacy through mechanisms like zero-knowledge proofs.

Succinct Non-Interactive Arguments of Knowledge (SNARKs) are widely utilized in argument systems \cite{gabizon2019plonk,chiesa2020marlin,gennaro2013quadratic,parno2016pinocchio,ben2014succinct,goldwasser2019knowledge}. They facilitate efficient verification of complex computations, such as those involving large circuits or datasets, by providing sublinear proof sizes and constant-time verification.

\begin{definition}[Zero-Knowledge Proof]
\label{def:zk}
An interactive proof system is zero-knowledge if there exists a PPT simulator $\mathcal{S}$ such that for all $(x,w) \in R$, and for any polynomial time distinguishers $\mathcal{D}$ (i.e., a polynomial time algorithm that attempts to distinguish between the real view of the verifier and the output of the simulator), we have:
\begin{equation}
\left| \begin{gathered}
  \Pr \left[ {\mathcal{D}\left( {{\text{Vie}}{{\text{w}}_\mathcal{V}}[\langle \mathcal{P}(x,w),\mathcal{V}(x)\rangle ]} \right) = 1} \right] \hfill \\
   - \Pr \left[ {\mathcal{D}(\mathcal{S}(x)) = 1} \right] \hfill \\ 
\end{gathered}  \right| \leqslant \mathsf{negl}(\lambda)    
\end{equation}
where $\text{View}_\mathcal{V}$ denotes the verifier’s view of the protocol execution.
\end{definition}

\subsubsection{Polynomial Commitments}
\label{def:polycommit}
A polynomial commitment scheme enables a prover to commit to a polynomial \( f \) of degree at most \( d \) over a finite field \(\mathbb{F}\), such that the prover can later open the evaluation \( y = f(x) \) at any point \( x \in \mathbb{F} \) with a succinct proof. Representative realizations include the Hyrax commitment~\cite{wahby2018doubly} and its multi-opening extensions~\cite{boneh2020efficient}. A commitment scheme must satisfy two essential security properties: 
\begin{itemize}
    \item \textbf{Binding}: No PPT adversary can produce a commitment \( c \) and later open it to two distinct values \( y \neq y' \) for the same point \( x \).
    \item \textbf{Hiding}: The commitment \( c \) reveals no information about the polynomial \( f \), beyond opened evaluations.
\end{itemize}

A polynomial commitment scheme usually consists of four algorithms: \(\mathsf{KeyGen}, \mathsf{Commit}, \mathsf{Open}, \mathsf{Verify}\), which are defined as follows:

\begin{itemize}
    \item \(\mathsf{pp} \leftarrow \mathsf{KeyGen}(1^\lambda, d)\): Generates public parameters \(\mathsf{pp}\) for polynomials of degree at most \( d \), given a security parameter \(\lambda\).
    \item \(c \leftarrow \mathsf{Commit}(\mathsf{pp}, f; r)\): Outputs a commitment \( c \) to a polynomial \( f \in \mathbb{F}_{\le d}[X] \) with randomness \( r \).
    \item \((y, \pi) \leftarrow \mathsf{Open}(\mathsf{pp}, f, x)\): Computes \( y = f(x) \) and generates a proof \(\pi\).
    \item \(1/0 \leftarrow \mathsf{Verify}(\mathsf{pp}, c, x, y, \pi)\): Checks if \( y \) is consistent with the commitment \( c \) at point \( x \), returning \( 1 \) if valid.
\end{itemize}

\subsection{Sumcheck, and MLE}

\subsubsection{Sumcheck Protocol}
The sumcheck protocol~\cite{cormode2012practical} is a fundamental component of modern interactive proof systems, facilitating verifiable computation of multivariate polynomial summations with sublinear complexity. Formally, given a polynomial \( f\colon \mathbb{F}^l \rightarrow \mathbb{F} \) of degree \( d \) in each variable, the protocol enables a verifier to confirm the correctness of a prover's claimed sum: $S = \sum_{b_1,\ldots,b_l\in\{0,1\}} f(b_1,\ldots,b_l)$ with communication and verification costs scaling logarithmically with size of \( 2^l \). 

The sumcheck protocol proceeds through \( l \) iterative rounds: in each round, the prover sends a univariate polynomial \( g_i(x_i) \), purportedly representing the partial sum of \( f \) over all remaining variables, while the verifier checks consistency by sampling a random challenge \( r_i \in \mathbb{F} \). After \( l \) rounds, the sumcheck protocol reduces the verification task to evaluating \( f \) at a single random point. The Schwartz-Zippel lemma~\cite{schwartz1980fast,zippel1979probabilistic} underpins this probabilistic check, confining the soundness error to \(\frac{d}{|\mathbb{F}|}\), where \( d \) is the degree of the polynomial and \(|\mathbb{F}|\) is the cardinality of the finite field.

However, the original sumcheck protocol is not zero-knowledge because the prover reveals evaluations of \( f \). References~\cite{xie2019libra,xie2022orion,chiesa2017zero} introduced zero-knowledge polynomial commitments to conceal \( f \) in the interactive proof procedure of sumcheck, enabling zero-knowledge variants. Additionally, the Fiat-Shamir transform can convert the protocol into a non-interactive zero-knowledge proof under the random oracle model.

\subsubsection{Multilinear Extensions}

Multilinear extensions (MLEs) are widely used in protocols like the sumcheck protocol, as they enable summations and evaluations to be extended to a larger domain. Specifically, given a Boolean function \( f\colon \{0,1\}^n \to \mathbb{F} \), its multilinear extension \( \widetilde{f}\colon \mathbb{F}^n \to \mathbb{F} \) is defined as:
\begin{equation}\label{eq:MLE}
\widetilde{f}(\mathbf{x}) = \sum_{\mathbf{b} \in \{0,1\}^n} f(\mathbf{b}) \cdot \prod_{i=1}^n \left(b_i x_i + (1 - b_i)(1 - x_i)\right)
\end{equation}
where $\mathbf{x}=(x_1,\ldots,x_n) \in \mathbb{F}^n$, $ \mathbf{b} = (b_1,\ldots,b_n) \in \{0,1\}^n$ are vectors over the finite filed and the binary filed, respectively. Note that each term $\prod\nolimits_{i = 1}^n {\left( {{b_i}{x_i} + (1 - {b_i})(1 - {x_i})} \right)} $ in (\ref{eq:MLE}) is a multilinear Lagrange basis polynomial that evaluates to \( 1 \) at \( \mathbf{x} = \mathbf{b} \) and \( 0 \) elsewhere for any $\mathbf{x} \in \{0,1\}^{n}$. This property makes \( \widetilde{f} \) easy to evaluate and manipulate over the entire field \( \mathbb{F} \), thereby bridging discrete computations with algebraic methods. 

In practice, MLEs are indispensable for modern cryptographic protocols. For instance, in the sumcheck protocol, MLEs transform discrete summations over \( \{0,1\}^n \) into polynomial evaluations over \( \mathbb{F}^n \), enabling the verifier to efficiently validate complex claims via interactive proofs. Similarly, zk-SNARK frameworks such as Spartan~\cite{setty2020spartan} leverage MLEs to compress circuit constraints into multilinear forms, significantly reducing both proof size and verification overhead.

\subsection{Verifiable Tensor Operation}
Tensor operations (i.e., the computations using matrices) serve as foundational components of LLMs \cite{radford2019language,achiam2023gpt,chowdhery2023palm,touvron2023llama}. When implemented in zero-knowledge proof systems, tensor operations fundamentally rely on the sumcheck protocol \cite{cormode2012practical} and MLEs \cite{lund1992algebraic}. Consider dimensional-two tensors (i.e., matrices) $A \in \mathbb{F}^{D_0 \times D_1}$, $B \in \mathbb{F}^{D_1 \times D_2}$, and $C \in \mathbb{F}^{D_0 \times D_2}$ that are defined over a finite field \( \mathbb{F} \), where \( C = A \cdot B \) represents the standard matrix product:
\begin{equation}
C[i,j] = \sum_{k=1}^{D_1 } A[i,k] \cdot B[k,j], \quad \forall i < D_0,\ j < D_2.
\label{eq:matmul}
\end{equation}
To standardize dimensionality, all tensor dimensions are padded to the nearest powers of two using zeros, i.e., \( D_0 = 2^{d_0}, D_1 = 2^{d_1}, D_2 = 2^{d_2} \), for some integers \( d_0, d_1, d_2 \). Let \( \widetilde{A} \), \( \widetilde{B} \), and \( \widetilde{C} \) denote the MLEs of \( A \), \( B \), and \( C \), respectively, allow the matrix multiplication property to be expressed succinctly. Specifically, the relationship is reformulated as a summation over binary index representations:
\begin{align}
\sum_{{w} \in \{0,1\}^{\log D_1}} 
\left( D_1^{-1} \cdot \widetilde{C}({u}, {v}) 
- \widetilde{A}({u}, {w}) \cdot \widetilde{B}({w}, {v}) \right) 
= 0
\label{eq:zk-mle-matmul}
\end{align}
where \( u \in \{0,1\}^{\log D_0} \), \( w \in \{0,1\}^{\log D_1} \), and \( v \in \{0,1\}^{\log D_2}\) represent the binary encodings of the matrix indices.

To verify this relationship expressed by (\ref{eq:zk-mle-matmul}), the sumcheck protocol operates interactively over a total of $d = d_0 + d_1 + d_2$ rounds. At the start, the prover claims that the following relation holds: $\sum_{i \in \{0,1\}^{d}} f(i) = 0$, where \(f(i) = D_1^{-1}\widetilde{C}(u,v) - \widetilde{A}(u,w)\widetilde{B}(w,v)\), and  $i = (u,w,v)$ concatenates the binary vectors $u, w, v$. In each round, the prover sends a univariate polynomial obtained by restricting one variable at a time, based on the verifier’s randomly chosen challenge. This process iteratively decomposes the summation into univariate polynomial claims, progressively reducing the problem’s dimensionality. After $d$ round, the verifier checks consistency by evaluating \( \widetilde{A} \), \( \widetilde{B} \), and \( \widetilde{C} \) at a random point \( \left(r_u, r_w, r_v\right) \in \mathbb{F}^d \), ensuring the correctness of the original tensor operation with soundness error  \( O\left(d / \left|\mathbb{F}\right|\right) \).

\subsection{Lookup Arguments for Non-arithmetic Operations}
In our work, we extend the lookup arguments \cite{sun2024zkllm}, as commonly employed to achieve non-arithmetic operations within the domain of zero-knowledge proofs~\cite{thaler2022proofs}, specially incorporate it into zero-knowledge verifiable deep learning inference~\cite{kang2022scaling} and LLM inference~\cite{sun2024zkllm}, to the domain of gradient computation in LLMs. Lookup arguments are cryptographic primitives designed to efficiently prove set-membership conditions, enabling the prover to demonstrate to a verifier the correctness of non-arithmetic operation in LLMs. Formally, the prover holds a secret table \( S = \{s_1,s_2,\ldots,s_D\} \subseteq \mathbb{F}^D \), where \( \mathbb{F} \) denotes a finite field. The lookup arguments enable a prover to prove to a verifier that each element \( s_i \) in $S$ is contained within a publicly known, predefined lookup table \( T = \{t_1,t_2,\ldots,t_N\} \subseteq \mathbb{F}^N \). 

Specifically, the set-membership condition \( S \subseteq T \) is satisfied if and only if there exists an auxiliary multiplicity tensor \( m = \{m_1,m_2,\ldots,m_N\} \subseteq \mathbb{F}^N \), such that the following rational-function identity holds:
\begin{align}\label{rational-function}
\frac{d}{dX} \log{\left( \prod_{i=1}^{D}(X + s_i) \right)} 
= 
\frac{d}{dX} \log{\left( \prod_{j=1}^{N}(X + t_j)^{m_j} \right)}
\end{align}
where each \( m_j \) indicates the multiplicity of the corresponding public table element \( t_j \) within the secret table \( S \): $m_j = |\{i:s_i=t_j\}|$ for $i\in\{1,2,\cdots,N\}$. According to zkLLM~\cite{sun2024zkllm}, the membership condition \( S \subseteq T \) can be equivalently represented in inner product form:
\begin{align}
\sum_{i \in [D]} A_i = \sum_{j \in [N]} m_j B_j
\end{align}
 where \( A_i = \frac{1}{\beta + s_i} \) represents the evaluation of an element \( s_i \) from the secret tensor \( S \) at a randomly chosen challenge \( X \gets \beta \sim \mathbb{F} \), and \( B_j = \frac{1}{\beta + t_j} \) represents the corresponding evaluation for an element \( t_j \) from the lookup table \( T \). Here, for any integer $n$, we denote $[n] = \{1, 2, \ldots, n\}$.

The right side of the equation, \( \sum_{j \in [N]} m_j B_j \), encodes a weighted version of the lookup table. To further improve efficiency and facilitate parallel computation, additional randomness elements \( \alpha \in \mathbb{F} \) and a binary vector \( u \in \mathbb{F}^{\log_2 D} \) are introduced in the equation. These elements enable the structuring of the verification process within a polynomial sumcheck framework, ensuring both succinctness and verifiability. The refined membership condition is then expressed as:
\begin{align}
\alpha + \alpha^2 =\ & 
\alpha \sum_{i,j} \widetilde{A}(i \oplus j)\, \widetilde{e}(u, i \oplus j)(S(i \oplus j) + \beta) \notag \\
& + \sum_{i \in [D/N]} \sum_{j \in [N]} \widetilde{A}(i \oplus j) \notag \\
& + ND^{-1} \alpha^2 \sum_{j \in [N]} \widetilde{B}(j)\, \widetilde{e}(u_{[\log_2(D/N):]}, j)(T(j) + \beta) \notag \\
& - ND^{-1} \sum_{j \in [N]} \widetilde{B}(j)\, \widetilde{m}(j)
\label{eq:lookup-sumcheck}
\end{align}

A complete description of the procedure to prove ${S} \subseteq  {T}$ is found in \textbf{Protocol \ref{protocol 1}}. In particular, in Line~1, $\mathsf{Setup}(T)$ generates a short witness $\llbracket T \rrbracket$ to a prescribed table $T$ known to both of the prover and the verifier; in Line~4, the prover  constructs ${m}$ based on ${S}$ and ${T}$ and commits to ${S}$ and ${m}$ using $\mathsf{Prep}({S}, {T})$; finally, in Line~9, $\langle \mathcal{P}, \mathcal{V} \rangle.\mathsf{Prove}(\llbracket {S} \rrbracket, \llbracket {m} \rrbracket, \llbracket T \rrbracket)$ is the interactive process between the prover and the verifier, proving that a secret tensor ${S}$ is element-wisely in $T$, which has been committed as $\llbracket T \rrbracket$.
\begin{algorithm}[!t]
\floatname{algorithm}{Protocol}
\caption{\textsc{Lookup Argument}}
\label{protocol 1}
\textbf{Require:} The prover $\mathcal{P}$ knows $S \in \mathbb{F}^D$, and the prover $\mathcal{P}$ and the verifier $\mathcal{V}$ both know  $T \in \mathbb{F}^N$, where the integers $N, D$ are both powers of $2$ such that $N$ divides $D$.
\begin{algorithmic}[1]

\Procedure{{$\mathsf{Setup}$}}{$T \in \mathbb{F}^N$}
    \State \Return $\llbracket T \rrbracket \gets \mathsf{Commit}(T)$ 
\EndProcedure

\Procedure{$\mathcal{P}.\mathsf{Prep}$}{$S \in \mathbb{F}^D, T \in \mathbb{F}^N$}
    \State Compute $m$ based on $S$ and $T$
    \State $\mathcal{P} \rightarrow \mathcal{V} : \llbracket S \rrbracket \gets \mathsf{Commit}(S)$
    \State $\mathcal{P} \rightarrow \mathcal{V} : \llbracket m \rrbracket \gets \mathsf{Commit}(m)$
\EndProcedure

\Procedure{$\langle \mathcal{P}, \mathcal{V} \rangle.\mathsf{Prove}$}{$\llbracket S \rrbracket, \llbracket m \rrbracket, \llbracket T \rrbracket$}
    \State $\mathcal{V} \rightarrow \mathcal{P} : \beta \sim \mathbb{F}$
    \State $\mathcal{P}$ computes $A$, $B$
    \State $\mathcal{P} \rightarrow \mathcal{V} : \llbracket A \rrbracket \gets \mathsf{Commit}(A), \llbracket B \rrbracket \gets \mathsf{Commit}(B)$
    \State $\mathcal{P}$ and $\mathcal{V}$ run sumcheck on \eqref{eq:lookup-sumcheck}, followed by the proofs of evaluation on $\llbracket A \rrbracket, \llbracket B \rrbracket, \llbracket S \rrbracket, \llbracket m \rrbracket, \llbracket T \rrbracket$
\EndProcedure

\end{algorithmic}

\end{algorithm}

\section{Gradient Computations for LoRA in LLMs}\label{llm}

This section begins by outlining the typical structure of a large language model (LLM), followed by an introduction to the concept and overall process of LoRA fine-tuning. Finally, it delves into the detailed gradient derivations involved in the LoRA fine-tuning process, which are crucial for determining how to apply ZKP to validate LoRA.
\begin{figure*}[!htbp]
  \centering
  \includegraphics[width=1\textwidth]{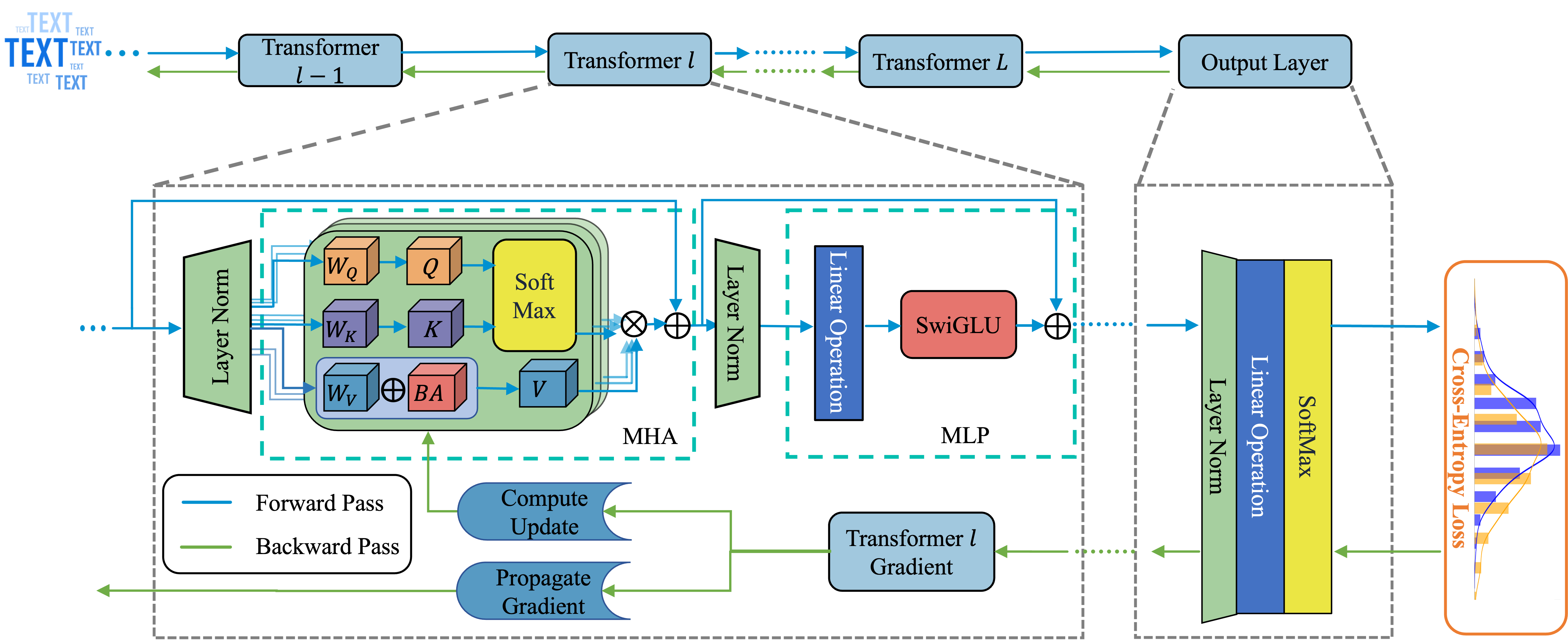}\\
  \caption{The typical structure of an LLM with Transformer layers, and the entire computation procedure of LoRA fine-tuning applied to an LLM. }\label{fig1.png}

\end{figure*}

\subsection{Structure of LLM}

A LLM, such as LLaMA~\cite{touvron2023llama}, is a specialized deep neural network architecture designed to map input token sequences to contextualized representations through a stack of Transformer layers. Fig.~\ref{fig1.png} illustrates a typical LLM architecture, which consists of $L$ cascaded Transformer layers followed by a final linear projection layer. In this subsection, we will introduce the specific structures and computations involved in the Transformer layers and the final output layer separately.

\subsubsection{Transformer Layers}

Each Transformer layer consists of two main components: a multi-head attention (MHA) sub-layer~\cite{vaswani2017attention} and a feed-forward multilayer perceptron (MLP) sub-layer. Let \( x \in \mathbb{R}^n \) denote the input token sequence to the LLM, and \( X_{\ell} \in \mathbb{R}^{n \times d} \), \(\ell = 1, \cdots, L\), denote the output of the \(\ell^{\text{th}}\) Transformer layer, where \( n \) represents the sequence length and \( d \) represents the hidden dimension. Since the $L$ Transformer layers are sequentially cascaded, \( X_{\ell-1} \) serves as the input to the \(\ell^{\text{th}}\) Transformer layer for \(\ell = 2, \cdots, L\). The input token sequence  \( x \)  is the input to the first Transformer layer. Below, we describe the structure and computations of the \(\ell^{\text{th}}\) Transformer layer, which takes \( X_{\ell-1} \) as input and computes \( X_{\ell} \) as output.

In LLaMA, each Transformer layer adopts a pre-normalization structure, where the input to each sub-layer is first normalized using LayerNorm before being processed by the sub-layer (i.e., the MHA or MLP sub-layer). Specifically, LayerNorm is applied row-wise to a matrix \( M \in \mathbb{R}^{n \times d} \), transforming it into a normalized matrix \( M' \in \mathbb{R}^{n \times d} \) as follows:
\begin{equation}
M' = \mathsf{LayerNorm}(M) \label{normal}
\end{equation}
such that:
\begin{equation}
M'_{i,j} = \frac{M_{i,j} - \mu_i}{\sqrt{\sigma_i^2 + \varepsilon}} \gamma_i + \beta_i
\end{equation}
where \( M_{i,j} \) (\( M'_{i,j} \)) represents the \((i, j)^{\text{th}}\) entry of \( M \) (\( M' \)), and \(\mu_i = \frac{1}{d} \sum_{j=1}^d M_{i,j}\) and \(\sigma_i^2 = \frac{1}{d} \sum_{j=1}^d (M_{i,j} - \mu_i)^2\) denote the mean and variance of the \(i^{\text{th}}\) row of \( M \), which corresponds to the representation of the \(i^{\text{th}}\) entry of the input token sequence \( x \). Here, \(\gamma_i\) and \(\beta_i\) are learnable scaling and bias parameters for the \(i^{\text{th}}\) row of \( M \), respectively. All \(\gamma_i\) values are collectively represented as a vector \(\gamma \in \mathbb{R}^d\), and all \(\beta_i\) values are represented as a vector \(\beta \in \mathbb{R}^d\). These vectors are distinguished based on the location of the LayerNorm operation, as follows: \(\gamma_{NormMHA,\ell}\): LayerNorm scaling vector before the MHA sub-layer of the \(\ell^{\text{th}}\) Transformer layer;  \(\gamma_{NormMLP,\ell}\): LayerNorm scaling vector before the MLP sub-layer of the \(\ell^{\text{th}}\) Transformer layer; \(\gamma_{NormOut}\): LayerNorm scaling vector before the final output layer.

Before the MHA sub-layer, LayerNorm is applied to normalize the input of this Transformer layer $X_{\ell-1}$:  $X'_{\ell-1}=\mathsf{LayerNorm}(X_{\ell-1})$. The output from LayerNorm, $X'_{\ell-1}$ is then passed through the  MHA sub-layer that computes its attention outputs as: 
\begin{equation}
\mathsf{MHA}(X'_{\ell-1}) = \sigma\left(\frac{Q_{\ell}K_{\ell}^\top}{\sqrt{d}}\right)V_{\ell} \label{att}
\end{equation}
where $\sigma(\cdot)$ is the softmax function; \( Q_{\ell} = X'_{\ell-1}W_{Q,\ell} \) is the query matrix,  \( K_{\ell} = X'_{\ell-1}W_{K,\ell} \) is the key matrix, and \( V_{\ell} = X'_{\ell-1}W_{V,\ell} \in \mathbb{R}^{d \times d} \) is the value matrix, which are  computed by using the query-projection, key-projection, and value-projection matrices \( W_{Q,\ell}, W_{K,\ell}, W_{V,\ell} \in \mathbb{R}^{d \times d} \), respectively. Here, the softmax function, $\sigma(\cdot)$, is applied in a row-wise manner to the input matrix $Q_{\ell}K_{\ell}^\top$, ensuring that each row is normalized into a valid probability distribution. The attention outputs from MHA are then concatenated and linearly transformed using a projection matrix $W_{P, \ell}$:
\begin{equation}
O_{\ell} = \mathsf{MHA}(X'_{\ell-1})W_{P,\ell}.
\label{transformer_o}
\end{equation}

Before the MLP sub-layer, LayerNorm is applied to normalize the residual-enhanced representation \( R_{\ell} = X_{\ell-1}+O_{\ell} \): $R'_{\ell} = \mathsf{LayerNorm}(R_{\ell})$. The output from LayerNorm $R^{'}_{\ell}$ is then passed through the feed-forward MLP with SwiGLU activation~\cite{touvron2023llama}: 
\begin{equation}
\begin{array}{l}
\mathsf{FFN}(R'_{\ell}) =   \left( \phi(R'_{\ell} W_{{gate}, \ell}^\top) \odot \right.  \left. R'_{\ell} W_{{up}, \ell}^\top \right)   W_{{down}, \ell}^\top
\end{array}
\label{transformer_ffn}
\end{equation}
where $\phi(\cdot)$ denotes the SwiGLU activation function,  \( W_{{up},\ell}, W_{{gate},\ell} \in \mathbb{R}^{h \times d} \) and \( W_{{down},\ell} \in \mathbb{R}^{d \times h} \), are the up-projection, gate-projection, and down-projection matrices, and \( \odot \) denotes element-wise product. The final output of this Transform layer is obtained by applying the residual-enhanced representation to the output of MLP: 
\begin{equation}
X_{\ell} = R_{\ell} + \mathsf{FFN}(R'_{\ell}).
\label{transformer_x}
\end{equation}

\subsubsection{Final Output Layer}
After the last Transformer layer, a final output layer is applied to map the output of the last Transformer layer to an estimate of the one-hot encoded label matrix corresponding to the input token sequence. 

Before the final output layer, the output of the last Transformer layer, $X_{L}$, is laso processed by LayerNorm to get $X'_{L}=\mathsf{LayerNorm}(X_{L})$. Then, the normalized one is projected to logits via a linear operation: $O_L = X'_{L}W_{Head}$, where $O_L \in \mathbb{R}^{n \times v}$ represents the logits, and $W_{Head} \in \mathbb{R}^{d \times v}$ denotes the linear projection matrix that maps the final hidden representations from the Transformer layers to the vocabulary space for token classification or generation. After that, the estimate of the one-hot encoded matrix $\hat{Y} \in \mathbb{R}^{n \times v}$ is obtained via applying the Softmax function on the logits, which can be expressed as $\hat{Y} =\sigma(O_L) $, where $\hat{Y}$ is the estimate of the one-hot encoded label matrix $Y \in \mathbb{R}^{n \times v}$. Each row of $Y$ corresponds to the one-hot encoding of the ground-truth token at the respective positions in the input sequence $x$, where $v$ denotes the vocabulary size of the training dataset.

\subsection{LoRA Fine Tuning}
\label{Sec:LoRA}

The fine-tuning of the LLM aims to optimize the parameters of the LLM, i.e., the projection matrices of MHA in all Transformer layers, the weight matrices of MLP in all Transformer layers and the final output layers, and the scaling and bias vectors in all LayerNorm operations of all layers. In the following, we focus on the projection and weight matrices of the $\ell^{\text{th}}$ Transformer layer: $\{W_{Q,\ell}, W_{K,\ell}, W_{V,\ell}, W_{{up},\ell}, W_{{down},\ell}, W_{{gate},\ell} \}$. In standard fine-tuning, the parameter matrices are directly updated during training, and fine-tuning LLMs is computationally expensive due to the large number of parameters (i.e., the high dimensions of these parameter matrices). 

Instead of directly updating the parameter matrices as in standard fine-tuning, LoRA~\cite{hu2022lora}  offers a parameter-efficient alternative by freezing the original parameter matrices and introducing trainable low-rank matrices into the architecture. These low-rank matrices are added to the original parameter matrices during both the forward and backward propagation phases of LLM fine-tuning. Mathematically, for a target matrix \( W_{\ell} \in \{W_{Q,\ell}, W_{K,\ell}, W_{V,\ell}, W_{{up},\ell}, W_{{down},\ell}, W_{{gate},\ell} \} \), LoRA modifies it in each iteration as:
$W'_{\ell} = W_{\ell} + \Delta W_{\ell}$ with $\Delta W_{\ell} = B_{\ell} A_{\ell}$. Here, $\Delta W_{\ell}$ represents the fine-tuning update, which is decomposed into two small low-rank matrices \( B_{\ell} \in \mathbb{R}^{d \times r} \) and \( A_{\ell} \in \mathbb{R}^{r \times k} \), with rank \( r \) satisfying \( r \ll \min{\left(d, k\right)} \). Since LoRA fine-tuning only the low-rank matrices \( B_{\ell} \) and \( A_{\ell} \), while keeping the original parameter matrices \( W_{\ell} \) frozen, the number of trainable parameters is significantly reduced compared to standard fine-tuning. This makes LoRA an efficient and scalable method for adapting large language models.

To further simplify the fine-tuning procedure, LoRA is typically applied to update one or some target parameter matrices rather than all the parameter matrices for each Transformer layer. Here, we describe the fine-tuning procedure of LoRA when it is specifically configured to update the value-projection matrices of the Transformer layers, $W_{V,\ell}$ for $\ell \in [L]$. For notational simplicity, we describe the fine-tuning procedure at the level of a single mini-batch for training epoch $t \in [T]$. Specifically, we denote the mini-batch in epoch $t$ as $\mathcal{B}^{t}$. Each data sample in $\mathcal{B}^{t}$ is represented by a tuple $(x, Y)$, where $x \in \mathbb{R}^{n}$ is an input token sequence of length $n$, and $Y \in \mathbb{R}^{n \times v}$ is a one-hot encoded label matrix. Each row of $Y$ corresponds to the one-hot encoding of the ground-truth token at the respective positions in the input sequence $x$, where $v$ denotes the vocabulary size of the training dataset. Fig.~\ref{fig1.png} illustrates the entire computation procedure of LoRA fine-tuning that consists of the following three phases per training epoch:

\begin{enumerate}
\item{Forward Propagation}: The token sequence $x$ are taken as the input to the first Transformer layer and propagated through each of all the Transformer layers sequentially to compute their outputs denoted by $X_{{\ell}} \in \mathbb{R}^{n \times d}$ for $\ell  = 1, \cdots ,L$. The final output layer computes the estimate on the one-hot encoded label matrix $\hat Y$ for $x$. Finally, the cross-entropy loss for epoch $t$ is computed as: 
\begin{equation}
\mathcal{L}^{t} = - \sum_{(x,Y) \in \mathcal{B}^{t} } \sum_{i=1}^{n} \sum_{j=1}^{v} y_{i,j} \log (\hat{y}_{i,j})\label{loss}
\end{equation}
where $y_{i,j}$ and $\hat{y}_{i,j}$ are the $(i,j)^{\text{th}}$ entry of the one-hot encoded label matrix $Y$ and its estimate $\hat{Y}$, respectively.

\item{Backward Propagation}: During backward propagation, the gradients of the parameters are computed recursively in reverse order, from the final layer to the first layer, based on the loss $\mathcal{L}^{t}$ using the chain rule. In LoRA fine-tuning,  only the value-projection matrix $W_{V, \ell}$ is updated, and it is replaced by 
\begin{equation}
W'_{V, \ell} = W_{V, \ell} + B_{\ell} A_{\ell}
\label{decompose}
\end{equation}
where the original projection matrices $W_{V, \ell}$ remain frozen for each $\ell$. The backward propagation in LoRA first computes the gradient of the modified value-projection matrix $\frac{\partial \mathcal{L}^{t}}{\partial W'^{t}_{V,\ell}}$, and subsequently computes the gradients of the low-rank matrices $A_\ell$ and $B_\ell$ for each Transformer layer $\ell$ using the chain rule: $\frac{\partial \mathcal{L}^{t}}{\partial B^{t}_{\ell}}=\frac{\partial \mathcal{L}^{t}}{\partial W'^{t}_{V,\ell}} \frac{\partial W'^{t}_{V, \ell}}{\partial B^{t}_{\ell}}$ and $\frac{\partial \mathcal{L}^{t}}{\partial A^{t}_{\ell}}=\frac{\partial \mathcal{L}^{t}}{\partial W'^{t}_{V,\ell}} \frac{\partial W'^{t}_{V,\ell}}{\partial A^{t}_{\ell}}$. We will present the detailed derivation of these gradients in the subsequent section.

\item{Parameter Update}:
With the computed gradients, the parameters targeted for updating at each Transformer layer, specifically the low-rank matrices $A_\ell$ and $B_\ell$, are updated using a learning rate $\eta$ as follows:
\begin{align}
\begin{gathered}
 A_\ell^{t+1} = A_\ell^{t} - \eta \frac{\partial \mathcal{L}^{t}}{\partial A^{t}_\ell} \hfill \\
  B_\ell^{t+1} = B_\ell^{t} - \eta   \frac{\partial \mathcal{L}^{t}}{\partial B^{t}_\ell} \hfill \\ 
\end{gathered} 
\label{eq:grad_update_ab}
\end{align}
which completes the parameter updating of Transformer layer $\ell$ in the training epoch $t$ of the LoRA fine-tuning. 

\end{enumerate}

\subsection{Gradient Computations}
In this subsection, we will provide a detailed derivation of the gradient computation process during backward propagation, as both the computation process and the resulting gradients are the objects to be proven in our zero-knowledge proof algorithm.

Without loss of generality, we focus on the gradient derivation for Transformer layer $\ell$ in training eppch $t$. Now, assume that the gradient of the loss function with respect to (wrt) the output of Transformer layer \(  \ell \), denoted as $\frac{\partial \mathcal{L}^{t}}{\partial X^{t}_{\ell}}$, has already been computed in Transformer layer \(  \ell +1 \) and propagated backward to Transformer layer \( l \). Within Transformer layer $\ell$, gradients propagate sequentially through MLP, LayerNorm, MHA, LayerNorm. The gradient wrt the output of MHA is computed as:
\begin{align}
\frac{\partial \mathcal{L}^{t}}{\partial O_\ell^{t}} 
    = & \ \frac{\partial \mathcal{L}^{t}}{\partial X_{\ell}^{t}} \cdot 
    \frac{\partial X_{\ell}^{t}}{\partial O_{\ell}^{t}} \notag \\
\label{eq:grad_ol}
\end{align}
where 
\begin{align}
\begin{gathered}
  \frac{{\partial X_\ell^{t}}}
{{\partial O_\ell^{t}}} = {W_{down,\ell}} \odot [{W_{up,\ell}} \odot \phi '(O_\ell^{t}{W_{gate,\ell}}){W_{gate,\ell}} \\ 
   + \phi (O_\ell^{t}{W_{gate,\ell}}){W_{up,\ell}}] \odot \gamma_{NormMLP,\ell}^ \top  \\ 
\end{gathered} 
\label{eq:grad_ox}
\end{align}
is computed using the relationships in (\ref{transformer_o})-(\ref{transformer_x}). Here, \( \phi'(\cdot) \) represents the derivative of the activation function in the MLP gating path. 

We proceed with back-propagating the computed gradient and applying the chain rule. The gradient wrt the value matrix $V_{\ell}^{t}$ is computed as 
\begin{equation}
\frac{{\partial {\mathcal{L}^{t}}}}
{{\partial V_\ell^{t}}} = \frac{{\partial {\mathcal{L}^{t}}}}
{{\partial O_\ell^{t}}}\frac{{\partial O_\ell^{t}}}
{{\partial V_\ell^{t}}}
\label{eq:grad_vl}
\end{equation}
where 
\begin{equation}
\frac{{\partial O_\ell^{t}}}
{{\partial V_\ell^{t}}} = {W_{P,L}} \sigma{\left( {\frac{{Q_\ell^{t}K_\ell^{t \top }}}
{{\sqrt {{d}} }}} \right)^ \top }
\label{eq:grad_vo}
\end{equation}
is computed from the relationship between the output of MAH and its value matrix: $O_\ell= \sigma\left( {{{{Q_\ell}K_\ell^ \top } \mathord{\left/
 {\vphantom {{{Q_\ell}K_\ell^ \top } {\sqrt d }}} \right.
 \kern-\nulldelimiterspace} {\sqrt d }}} \right){V_\ell}{W_{P,\ell}}$.

In LoRA, $V_{\ell}$ can be expressed as $V_{\ell} = \mathsf{LayerNorm}(X_{\ell - 1}){W'_{V,\ell}} =X'_{\ell}{W'_{V,\ell}}$. Therefore, the gradient wrt the modified value-projection matrix $W'_{V,\ell}$ is computed as 
\begin{align}
\frac{\partial \mathcal{L}^{t}}{\partial W'^{t}_{V,\ell}} = \frac{{\partial {\mathcal{L}^{t}}}}
{{\partial V_\ell^{t}}} \frac{{\partial { V_\ell^{t}  }}} 
{{\partial W'^{t}_{V,\ell}}} = \frac{{\partial {\mathcal{L}^{t}}}}
{{\partial V_\ell^{t}}} X'^{ \top}_{\ell}
\label{eq:grad_wl}
\end{align}
Using the relationship in (\ref{decompose}), the gradients wrt the trainable low-rank matrices $B_\ell$ and $A_\ell$ are finally computed as:
\begin{align}
\begin{gathered}
  \frac{\partial \mathcal{L}^{t}}{\partial B^{t}_{\ell}}=\frac{\partial \mathcal{L}^{t}}{\partial W'^{t}_{V,\ell}} \frac{\partial W'^{t}_{V, \ell}}{\partial B^{t}_{\ell}}=\frac{\partial \mathcal{L}^{t}}{\partial W'^{t}_{V,\ell}} A_{\ell}^{t \top} \hfill \\
  \frac{\partial \mathcal{L}^{t}}{\partial A^{t}_{\ell}}=\frac{\partial \mathcal{L}^{t}}{\partial W'^{t}_{V,\ell}} \frac{\partial W'^{t}_{V, \ell}}{\partial A^{t}_{\ell}}=\frac{\partial \mathcal{L}^{t}}{\partial W'^{t}_{V,\ell}} B_{\ell}^{t \top}\hfill \\ 
\end{gathered} 
\label{eq:grad_abl}
\end{align}

In the following sections, we will demonstrate how our proposed VeriLoRA achieves verifiable computations throughout the entire LoRA procedure using zero-knowledge proofs.

\section{Design of VeriLoRA}
\label{sec:VeriLoRA}

The goal of VeriLoRA is to introduce zero-knowledge verifiability into LoRA fine-tuning for LLMs. In VeriLoRA, a prover is responsible for performing the computation required for LoRA fine-tuning on the LLM and generating cryptographic proofs that ensure all computational steps strictly adhere to the predefined procedures without deviation. Meanwhile, a verifier can efficiently validate these proofs without accessing any privacy-sensitive information, such as model parameters or training data. Owing to the zero-knowledge nature of the proofs, the verifier can confirm whether the LoRA fine-tuning process was executed correctly, without extracting any additional sensitive details.

The computations involved in LoRA fine-tuning can be broadly categorized into two types: arithmetic operations and non-arithmetic operations. Arithmetic operations, such as matrix additions and multiplications are inherently compatible with most mainstream ZKP systems. In contrast, non-arithmetic operations cannot be directly processed by the mainstream ZKP systems. As a result, specialized ZKP-compatible constraint systems are necessary to encode the computational logic of these non-arithmetic operations, particularly during the backward propagation phase of LoRA fine-tuning, which involves gradient computations. Therefore, in the following discussions, we will primarily focus on the proving protocols for these non-arithmetic operations.

As discussed in Section~\ref{Sec:LoRA}, the entire computation procedure of LoRA can be divided into three sequential phases, and VeriLoRA follows the same structure. In Section \ref{sec:zk_lora}, we detail the key technical aspects of the backward propagation phase in VeriLoRA as a representative example to illustrate how non-arithmetic operations are proven. The proving scheme for non-arithmetic operations in the forward propagation and parameter update phases follows the same approach. In Section \ref{sec:everything}, we will put everything together to present the complete VeriLoRA framework.


\subsection{VeriLoRA for Backward Propagation Phase}
\label{sec:zk_lora}
During the backward propagation phase, the gradients of the cross-entropy loss \( \mathcal{L}^{t} \) wrt the parameters are computed and recursively propagated backward through each Transformer layer \( \ell \), starting from the final output layer. For the backward propagation phase at each Transformer layer, we essentially aim to prove the gradient computation process expressed by (\ref{eq:grad_ol})-(\ref{eq:grad_abl}). The arithmetic operations in the backward propagation phase such as matrix additions and multiplications, are the same as the forward propagation phase, and we can encode the relevant matrices as MLEs and then prove the computation correctness using the sumcheck protocol, as proposed in \cite{sun2024zkdl,sun2024zkllm}. In contrast to the forward propagation phase, we face increased complexity in the backward propagation phase due to the gradient computations, which involve more non-arithmetic functions.

Specifically, the computations of the gradients in (\ref{eq:grad_ol})-(\ref{eq:grad_abl}) encompass the following non-arithmetic operation: the SwiGLU activation $\phi(\cdot)$ and its derivative $\phi'(\cdot)$, the Softmax operation $\sigma(\cdot)$, the matrix transposition $(\cdot)^{\top}$, and the element-wise product between two matrices $\odot$. To prove these non-arithmetic operations using ZKP, we design a dedicated proof protocol for each operation, primarily leveraging lookup-based arguments in conjunction with the sumcheck protocol. In the following, we present the details of the protocols for proving each of these non-arithmetic operations.

\subsubsection{\bf zkElementProd}
Using Protocol \ref{protocol 1}, we develop the zkElementProd protocol to prove the correctness of the computation of an element-wise product between two matrices: \( C = A \odot B \), where \( A, B, C \in \mathbb{F}^D \). The prover constructs the compressed secret set \( S \coloneqq \{C_i + \alpha \cdot (A_i \odot B_i)\}_{i=1}^D \), where \( \alpha \in \mathbb{F} \) is a random verifier challenge. The verifier constructs a public lookup table \( T \coloneqq \{c + \alpha \cdot a \cdot b \mid a, b \in \mathbb{F},\ c = a \cdot b\} \), where \( a, b \in \mathbb{F} \) range over all possible input values for the matrices \( A \) and \( B \), respectively, and \( c \) defines the corresponding output. The prover then proves the multiset inclusion \( S \subseteq T \) using Protocol \ref{protocol 1}, thereby ensuring that the element-wise product relation holds over the entire matrix.

\subsubsection{\bf zkTranspose}
We then develop the zkTranspose protocol to prove that a matrix \( A' \in \mathbb{F}^{d \times n} \) is indeed the transpose of a matrix \( A \in \mathbb{F}^{n \times d} \), i.e., \( A' = A^\top \). We denote the \((i,j)^{\text{th}}\) element of \( A \) by \( a_{i,j} \), and the \((i,j)^{\text{th}}\) element of \( A' \) by \( a'_{i,j} \). Given a random scalar \( \alpha \in \mathbb{F} \) chosen by the verifier, the prover constructs a compressed secret set \( S \) as: $S = \{a'_{i,j} + \alpha \cdot a_{j,i} \}$,  where $i \in [d],~j \in [n]$, and assigns the publicly known set as: $T = \{x + \alpha \cdot x\}$, where $x \in \mathbb{F}_d$. Here, in $T$, each $x$ represents a possible value of the elements in the matrix $A$ input to the transpose operation. To prove the correctness of the transpose operation, we apply Protocol \ref{protocol 1} to demonstrate that the set-membership condition \( S \subseteq T \) is satisfied.


\subsubsection{\bf zkSwiGLU}
We first establish the zkSwiGLU protocol, which achieves a zero-knowledge proof for the computations of the SwiGLU activation $\phi(\cdot)$ and its derivative $\phi'(\cdot)$. Note that both the SwiGLU activation $\phi(\cdot)$ and its derivative $\phi'(\cdot)$ apply a non-linear mapping to their input matrices in an element-wise manner. Since the proof protocols for $\phi(\cdot)$ and $\phi'(\cdot)$ are identical in VeriLoRA, we present the details of the proof protocol for $\phi(\cdot)$ as a representative.

We introduce the following notations for the presentation of the zkSwiGLU protocol. Let $B \in \mathbb{N}$ be a positive integer, and let $\mathbb{F}_B \coloneqq \{x \in \mathbb{F} : -\frac{B}{2} \leq x < \frac{B}{2} - 1\}$ denote a bounded domain. Let $Z$ be the matrix input to the activation function $\phi(\cdot)$, and $G=\phi(Z)$ be the corresponding matrix output from $\phi(\cdot)$. Let \( D \) denote the number of elements in the matrix \( Z \) (or \( G \)). Each element of \( Z \) (or \( G \)) is represented as \( z_i \) (or \( g_i \)), where \( g_i = \phi(z_i) \), and \( i \in [D] \). Finally, let $\zeta \in \mathbb{N}$ denote a scaling factor for the input matrix $Z$, chosen such that all scaled elements fall within the range of $\mathbb{F}_B$.

In zkSwiGLU, we decompose each of the elements in input matrix $Z$ as $z_{i}\coloneqq\zeta z_{i}'+r_i$, where $z_{i}'\coloneqq \left\lfloor z_{i}/\zeta \right\rfloor \in \mathbb{F}_B$ is the quantized element lying in the bounded domain, and $r_i\in [-\zeta/2, \zeta/2)$ is the residue after quantization. Here, $\left\lfloor \cdot \right\rfloor$ denotes the floor operator. To prove the computation correctness of the SwiGLU activation $G=\phi(Z)$, the prover of zkSwiGLU applies Protocol \ref{protocol 1} separately to prove each of the following lookup arguments such that each element of a secret set \( S \) belongs to a predefined set \( T \) without explicitly revealing the elements of \( S \):

\begin{enumerate}
\item \textbf{Activation Correctness}:  The secret set \( S \) is designated as \( S = \{z'_i + \alpha g_i\}_{i=1}^D \), where the random linear combination coefficient \( \alpha \in \mathbb{F} \) is chosen by the verifier \footnote{To ensure the protocol is non-interactive, modern cryptographic protocols of this type typically leverage the Fiat-Shamir heuristic, which eliminates the need for direct interaction between the prover and verifier. Specifically, each verifier challenge is computed by hashing all previously generated commitments and public messages in the protocol, making the protocol non-interactive.}. The predefined set \( T \) is assigned as \( T = \{x + \alpha \phi(x) \mid x \in \mathbb{F}_B\} \). Here, in \( T \), each \( x \) represents a possible value of the quantized activation input \( z'_i \), and \( \phi(x) \) denotes the corresponding activation output. The satisfaction of this lookup argument ensures the correctness of the computation relationship between the activation function's input and output, rather than any other arbitrary computations.

\item \textbf{Quantization Correctness}: 
The secret set \( S \) is designated as \( S = \{\zeta z'_i + r_i\}_{i=1}^D \), while the predefined set \( T \) is assigned as \( T = \{\zeta x + r \mid x \in \mathbb{F}_B,  r\in[-\zeta/2, \zeta/2)\} \). Here, in \( T \),  $x \in \mathbb{F}_B$ corresponds to a possible value of the quantized input element to the activation function (i.e., the possible value of  $z'_i$), and $r\in[-\zeta/2, \zeta/2)$ corresponds to a possible value of the quantization residue $r_i$. The satisfaction of this lookup argument ensures the correctness of the quantization process from the original value \( z_i \) to its quantized value \( z'_i \), for $i \in [D]$.

\item \textbf{Residue Correctness}: The secret set \( S \) is designated as \( S = \{r_i \}_{i=1}^D \). The predefined set \( T \) is assigned as $T=\{-\frac{\zeta}{2},\ldots,\frac{\zeta}{2}-1\}$.  The satisfaction of this lookup argument ensures that each quantization residue from the quantization process lies within the validate range. 
    
\end{enumerate}

The three lookup arguments described above achieve the zero-knowledge proof for the computations of the SwiGLU activation function \( S=\phi(G) \). Similarly, the zero-knowledge proof for the computations of the derivatives of the SwiGLU activation follows the same procedure, with the only difference being the replacement of the relationship between the input and output matrices to \( S = \phi'(G) \).

\subsubsection{\bf zkSoftmax}

We proceed to establish the zkSoftmax protocol, which offers a zero-knowledge proof for the computations involved in the Softmax operation \( \sigma(\cdot) \). The Softmax operation is applied row-wise to the input matrix, transforming each row of the input matrix into a corresponding row in the output matrix. Therefore, we construct the proof for each row of the Softmax input and output: \( p = \sigma(z) \), where \( z = [z_1, z_2, \cdots, z_n] \in \mathbb{F}^n \) and \( p = [p_1, p_2, \cdots, p_n] \in [0,1]^n \) represent the input and output vectors of length \( n \). Here, \( z_i \) and \( p_i \) denote the \( i^{\text{th}} \) elements of \( z \) and \( p \), respectively. The Softmax operation is mathematically expressed as: $p_i = \frac{e^{z_i/\xi}}{\sum_{j=1}^{n} e^{z_j/\xi}} \quad \text{for } i \in [n]$, where \( \xi \in \mathbb{N} \) is a fixed scaling factor chosen to control the dynamic range of exponent inputs.

The Softmax operation includes exponentiation and normalization, which are non-arithmetic and incompatible with arithmetic circuits. To address the multivariate and highly non-arithmetic nature of the Softmax operation, we utilize the shift-invariance property of Softmax, which states that adding a constant offset to all entries of the input vector does not change the output distribution. We further exploits Protocol \ref{protocol 1} to support proving the correctness of the computation of Softmax. Specifically, let the normalization shift be defined as: 
${\xi'} \coloneqq \xi \ln\left( \sum_{j=1}^{n} e^{z_j/\xi} \right)$. The Softmax output for computing the \( i^{\text{th}} \) output element can then be equivalently expressed in a shift-invariant form: $p_i = e^{(z_i-\xi')/\xi}$,
which follows from \( \sum_{j=1}^{n} e^{(z_j-\xi')/\xi} = 1 \). Next, the shifted input to the exponentiation function, \( \frac{z_i - {\xi'}}{\xi} \), is quantized to: $x_i \coloneqq \left\lceil\frac{z_i - {\xi'}}{\xi}\right\rceil \in (-B,0] \subset \mathbb{F}$, where \( B \) is a positive integer controlling the quantization granularity, and \( \left\lceil \cdot \right\rceil \) is the ceiling operator. Then, using \( K \) predetermined radices, \( \{b^{(k)}\}_{k=0}^{K-1} \), each quantized input \( x_i \) is decomposed into a product of \( K \) positive integers, as follows:
\begin{align}\label{decom_x}
x_i = &-\sum_{k=0}^{K} \left(\prod_{j=0}^{k-1} b^{(j)}\right)x_i^{(k)} 
\end{align}
where each digit \( x_i^{(k)} \in \{0,\ldots,b^{(k)}-1\} \) is bounded by its respective radix \( b^{(k)} \). For example, a set of predetermined radices could be \( K=2 \), \( b^{(0)}=2^{8} \), and \( b^{(1)}=2^{20} \). Consequently, the Softmax operation that computes the \( i^{\text{th}} \) output \( p_i \) can be decomposed into as the product of $K$ exponentiations: 
\begin{align}\label{decom}
  p_i = e^{x_i} = \prod_{k=1}^{K} e^{-\left(\prod_{j=1}^{k} b^{(j)}\right)x_i^{(k)}}=\prod_{k=1}^{K} y_i^{(k)}
\end{align}
where $y_i^{(k)}=e^{-\left(\prod_{j=1}^{k} b^{(j)}\right)x_i^{(k)}}$ is defined as the $k^{\text{th}}$ exponentiation. In the following, we will construct \( K \) public tables such that each table defines the input-output relationship of the \( k^{\text{th}} \) exponentiation in \eqref{decom}, and exploit lookup arguments on these tables to prove the correctness of \eqref{decom}. Compared to using a single table to define the input-output relationship of \( p_i = e^{x_i} \), this decomposition of \( p_i = e^{x_i} \) and the use of \( K \) tables for the lookup argument can significantly reduce the proving complexity, especially when the range of \( x_i \) is large.



Specifically, to prove the correctness of the computation in \eqref{decom} for all \( i \in [n] \) (i.e., to prove the Softmax computation \( p_i = \frac{e^{z_i/\xi}}{\sum_{j=1}^{n} e^{z_j/\xi}} \)), the prover needs to demonstrate that the following three conditions are satisfied:
\begin{itemize}
    \item \textbf{Decomposition Correctness}: The prover demonstrates that each $x_i$ is correctly decomposed to the digits on the $K$ radices, satisfying the condition specified in (\ref{decom_x}). This is achieved through the sumcheck protocol, where $x_i$ and $\{x_i^{(k)}\}_{k=0}^{K}$ are designated as private inputs, while \( \{b^{(k)}\}_{k=0}^{K-1} \) and the commitments of $x_i$ and $\{x_i^{(k)}\}_{k=0}^{K}$ are designated as public inputs. 

    \item \textbf{Exponentiation Correctness}: We apply Protocol \ref{protocol 1} to prove the exponentiation on each radix level, $y_i^{(k)}=e^{-\left(\prod_{j=1}^{k} b^{(j)}\right)x_i^{(k)}}$, is correct. For each radix level \( k \in [K] \), we define a public lookup table $T^{(k)} = \{(x, y): x, y = \left\lfloor e^{-(\prod_{j=1}^{k} b^{(j)})x}\right\rfloor \mid x \in \{0,\ldots,b^{(k)}-1\}\}$, which enumerates all valid input-output pairs to the \( k^{\text{th}} \) exponentiation in \eqref{decom}. Here, in $T^{(k)}$, $ x \in \{0,\ldots,b^{(k)}-1\}\}$ corresponds to a possible value of the decomposed digit $x_i^{(k)}$, and $y$ represents its corresponding exponentiation $y_i^{(k)}$. The secret table, $S^{(k)}$, just contains a single pair of $(x_i^{(k)}, y_i^{(k)})$. The satisfaction of the lookup argument, $S^{(k)} \subseteq T^{(k)}$, ensures the correctness of the exponentiation process at each digit level from $x^{(k)}$ to $y^{(k)}$, for $k \in [K]$.


    \item \textbf{Product Correctness}: Finally, the prover demonstrates that the full exponentiation value $p_i$ is correctly reconstructed as the product of $K$ exponentiations: $p_i = \prod_{k=1}^{K} y_i^{(k)}$ by using the zkElementProd protocol. 
\end{itemize}
The combination of the above sumcheck, lookup argument, and zkElementProd protocols collectively achieves a zero-knowledge proof for the Softmax operation.



\begin{algorithm}[!htbp]
\floatname{algorithm}{Protocol}

\caption{VeriLoRA: Zero-Knowledge Proof Framework for Verifiable LoRA Fine-tuning on LLMs}
\label{alg:VeriLoRA-final}
\begin{algorithmic}[1]

\State \textbf{Input}: The transformer model with pre-trained parameters that remain frozen during the fine-tuning; the randomly initialized low-rank decomposition matrices $A_{\ell}^{1}$ and $B_{\ell}^{1}$ for the value projection weights $W_{V,\ell}$ of all Transformer layers, $\ell=[L]$, the learning rate $\eta$, and the training data set. 

\vspace{0.1cm}\hrule\vspace{0.1cm}
\For{$t = 1$ to $T$}

  \begin{itemize}
      \item Sample the mini-batch $\mathcal{B}^{t}$ for training iteration $t$ from the training dataset;
      \item Take data $(x, Y)$ from $\mathcal{B}^{t}$ and input token sequence $x$   to the LLM; 
  \end{itemize}
\State  \textbf{Phase 1: Forward Propagation} 
\For{$\ell = 1$ to $L$}

\begin{enumerate}
\item[ ]
\begin{itemize}
      \item  Compute the output of each Transformer layer $\ell$, as \eqref{att}-\eqref{transformer_x}, and prove the related computations using \textbf{zkMat}, \textbf{zkTranspose}, \textbf{zkElementProd}, \textbf{zkSoftmax} and \textbf{zkSwiGLU}, respectively.
\end{itemize}
\end{enumerate}     
\EndFor
\begin{itemize}
    \item Compute the final output of the LLM, $\hat{Y}$, and the corresponding cross-entropy loss $\mathcal{L}^{t}$ using the one-hot encoded label matrix $Y$; 
\end{itemize} 
    
\State\textbf{Phase 2: Backward Propagation}
\For{$\ell = L$ to $1$}

\begin{enumerate}
\item[ ]
    \begin{itemize}
    \item Computes the gradients through each Transformer layer $\ell$, as \eqref{eq:grad_ol}-\eqref{eq:grad_abl}, and prove the related computations using \textbf{zkMat}, \textbf{zkTranspose}, \textbf{zkElementProd}, \textbf{zkSoftmax} and \textbf{zkSwiGLU}, respectively.
\end{itemize}

\end{enumerate}

\EndFor

\State \textbf{Phase 3: Parameter Update}
\begin{enumerate}
\item[ ]
\begin{itemize}
    \item Computes the parameter updates as in \eqref{eq:grad_update_ab} for all $\ell \in [L]$, and prove these computations using \textbf{zkMat}.
\end{itemize}   
\end{enumerate}
\EndFor


\State \textbf{Output}: The updated low-rank matrices $A_{\ell}^{T}$ and $B_{\ell}^{T}$ for all Transformer layers, $\ell \in [L]$. A set of proofs attesting to the correctness of all computations across Phases 1–3 for $T$ iterations.

\end{algorithmic}

\end{algorithm}

\subsection{putting everything together}
\label{sec:everything}
We formally present the complete {VeriLoRA} protocol, a zero-knowledge proof framework designed specifically to enable verifiable LoRA fine-tuning on Transformer based LLMs. {VeriLoRA} rigorously proves correctness at each computational step of LoRA fine-tuning, encompassing the forward and backward propagations through the Transformer layers, as well as the parameter update.

To achieve this, we employ sumcheck based protocol to prove all the matrix additions and multiplications operations that appear during the forward propagation, backward propagation, and parameter update phases. We call this protocol for proving linear matrix operations as \textbf{zkMat}. The \textbf{zkMat} protocol leverages the sumcheck-based veriable matrix operations described in Section \ref{sec:Preliminary} to prove all matrix multiplications and additions in LoRA. Given input matrices and their claimed output results, \textbf{zkMat} converts them to MLEs and applies the sumcheck protocol to verify the summations over MLEs. The zero-knowledge property is achieved through polynomial commitments to the MLEs, ensuring that the actual matrix data remains hidden while enabling verification. 

In contrast, all non-arithmetic operations that cannot be directly convert to arithmetic circuits are verified using our specially designed protocols described above, which are built upon lookup arguments, specifically: \textbf{zkSwiGLU} for proving the correctness of SwiGLU activation and its gradient computation in the MLP block; \textbf{zkSoftmax} for proving the correctness of Softmax function in self-attention mechanism; \textbf{zkElementProd} for proving the correctness of element-wise products and \textbf{zkTranspose} for proving the correctness of transposition operations during the forward and backward propagation of the self-attention mechanism. 

To integrate these protocols for proving different operations of the VeriLoRA fine tuning on LLM, we follow the same sequential-composition strategy as zkLLM \cite{sun2024zkllm}: i) Fixing random point-value claims through sumcheck/lookup, thereby binding the prover to specific evaluation points and claimed values; ii) Requiring each subsequent proving step to justify exactly the same points via further randomized reductions (e.g., random aggregation), conditioned on success in prior protocols; iii) Finalizing by opening evaluations against the same committed parameters, leveraging the binding property of the commitment scheme. These steps prevent the prover from reordering the proving steps or retroactively altering intermediate values, as both challenges and commitments are binding and fresh across proving steps.

Protocol~\ref{alg:VeriLoRA-final} presents {VeriLoRA} as a pseudo code, integrating the above protocols into a complete proving procedure for LoRA fine turning; subproofs are assembled in the arithmetic circuit’s reverse logical order.

\section{Security Analysis}
\label{sec:Security Analysis}
In this section, we present a formal security analysis of the VeriLoRA protocol, focusing on its completeness, soundness, and zero-knowledge properties under standard cryptographic assumptions. 

The VeriLoRA protocol is designed and implemented using the Hyrax polynomial commitment scheme, whose security guarantees are thoroughly discussed in \cite{raffel2020exploring}. To achieve non-interactive proof functionality, the protocol leverages the Fiat–Shamir heuristic, which operates under the standard model of a random oracle. This approach aligns with the formal frameworks commonly adopted in non-interactive proof systems \cite{raffel2020exploring}. Additionally, we assume the use of a sufficiently large finite field $\mathbb{F}$, typically with $|\mathbb{F}| \approx 2^{254}$, such as that derived from the BLS12-381 elliptic curve.


\subsection{Soundness}

\textbf{Statement}: VeriLoRA ensures soundness, meaning that a verifier will accept proofs for incorrectly computed operations from a malicious PPT prover who deviates from the computations specified in Protocol~\ref{alg:VeriLoRA-final} only with negligible probability. This probability is bounded (using a union bound with conditional probabilities) by: $\varepsilon_{\mathrm{sound}} \le \sum_{j}\varepsilon_j$, where $\varepsilon_j$ represents the soundness error of each proving step, conditioned on the correctness of all prior steps. Each $\varepsilon_j$ is defined by the following values:  $\frac{m d_{\mathrm{max}}}{|\mathbb{F}|}$ (for arithmetic operations),  $\frac{C}{|\mathbb{F}|}$ (for non-arithmetic operations), and   $\varepsilon_{\mathrm{binding}}$ (for polynomial commitments). Here, $m$ denotes the number of iteration rounds in the sumcheck protocol, $d_{\mathrm{max}}$ is the largest degree with respect to any single variable among all polynomials involved in the sumcheck protocol, $C$ represents the number of elements verified by a lookup argument in a non-arithmetic operation (e.g., element-wise products, SwiGLU activations and their gradient computations, Softmax operations, and matrix transpositions), and $\varepsilon_{\mathrm{binding}}$ refers to the binding error probability of Hyrax-style polynomial commitments.

\textbf{Analysis}: The soundness property of VeriLoRA is upheld by the protocol's sequential composition strategy that prevents the prover from reordering proving steps or retroactively modifying intermediate values, as both challenges and commitments are binding and freshly generated at each proving step. Consequently, the overall soundness error of VeriLoRA is bounded by the union bound with conditional probabilities (as given above in the statement), which represents the standard upper bound for sequential composition without assuming independence between proving steps. We analyze each $\varepsilon_j$, which represents the soundness error of VeriLoRA at each proving step, for arithmetic computations, non-arithmetic computations, and polynomial commitments, respectively.
\begin{itemize}
    \item \textbf{Arithmetic Operations}: VeriLoRA proves the arithmetic operations in LoRA (specifically, the matrix multiplications and additions) by representing them as MLEs and leveraging the sumcheck protocol for proof generation. According to the Schwartz-Zippel lemma and the analysis result in \cite{thaler2022proofs}, a false summation result over the polynomial is accepted with probability at most $m d_{\mathrm{max}}/|\mathbb{F}|$.

    \item \textbf{Non-arithmetic Operations}: VeriLoRA proves non-arithmetic computations in LoRA (e.g., element-wise products, SwiGLU activations, Softmax operations, and matrix transposes) using lookup-based arguments and sumcheck protocols. The prover employs the sumcheck protocol to verify rational-function identities, ensuring set-membership conditions ($S \subseteq T$) are satisfied. Specifically, the sum $\sum_{i=1}^D \frac{1}{X + s_i}$ equals $\sum_{j=1}^N \frac{m_j}{X + t_j}$, where $s_i$ are elements of the secret tensor $S$, $t_j$ are elements of the lookup table $T$, $m_j$ is the multiplicity of $t_j$ in $S$, and $X$ is a random challenge from $\mathbb{F}$. Any mismatch is detected with probability at most $C/|\mathbb{F}|$, ensuring negligible success for malicious provers.

    
    \item \textbf{Polynomial Commitments}: The Hyrax-style polynomial commitments utilized in VeriLoRA are computationally binding, ensuring that once a polynomial is committed, the prover cannot open it to a different value except with negligible probability $\varepsilon_{\mathrm{binding}}$. A violation of the binding property would constitute a breach of the discrete logarithm assumption in the random oracle model \cite{wahby2018doubly}, which is considered infeasible in practice.
\end{itemize}

In summary, VeriLoRA ensures that the soundness error for any invalid computation is bounded by the sum of the these error components described above. As a result, the overall soundness error remains negligible, guaranteeing that a malicious prover has an exceptionally low probability of persuading the verifier to accept an incorrect computation.

\subsection{Completeness}

\textbf{Statement}: VeriLoRA ensures completeness, meaning that a verifier will always accept proofs generated by an honest PPT prover who accurately performs all computations as specified in the VeriLoRA protocol described in Protocol~\ref{alg:VeriLoRA-final}.  

\textbf{Analysis}: We analyze the completeness of VeriLoRA in its arithmetic computations and non-arithmetic computations, respectively.

\begin{itemize}
    \item \textbf{Arithmetic Operations}: VeriLoRA proves the arithmetic computations in LoRA using the sumcheck protocol, which is known to have perfect completeness. This implies that when the prover is honest and performs all computations correctly, the verifier will accept the proof with probability exactly one.

    \item \textbf{Non-arithmetic Operations}: VeriLoRA proves the non-arithmetic computations in LoRA by employing lookup-based arguments. Completeness may fail if the random challenge from the verifier (e.g., $\beta$) coincides with a pole of the rational function used in the identity check of lookup-based arguments ~\cite{sun2024zkllm}. However, the probability of such an event is bounded by $ \frac{C}{|\mathbb{F}|}$, which is negligible for a sufficiently large field size.
\end{itemize}

In summary, VeriLoRA ensures that the completeness error is negligible, guaranteeing that an honest prover who performs all computations correctly will have an overwhelming probability of convincing the verifier to accept the proof.

\subsection{Zero Knowledge}

\textbf{Statement}: VeriLoRA ensures zero knowledge, meaning that the verifier learns nothing beyond the validity of the statement being proven. Formally, there exists a polynomial-time simulator $\mathcal{S}$ capable of generating transcripts indistinguishable from real interactions, even without access to the prover’s secret witness. The zero-knowledge error is negligible, bounded by $\mathsf{negl}(\lambda)$ under standard cryptographic assumptions.

\textbf{Analysis}: The zero-knowledge property of VeriLoRA is achieved by incorporating the Hyrax-style polynomial commitments~\cite{raffel2020exploring} into the sumcheck protocol~\cite{sun2024zkllm}, which provides computational hiding under the discrete logarithm assumption in the random oracle model. Specifically, VeriLoRA ensures that the verifier learns nothing beyond the validity of the computation while keeping private: 1) the pre-trained model parameters, 2) the low-rank matrices and their updates, 3) the internal computation values in SwiGLU, Softmax, transpose, and element-wise operations during the forward and backward propagations, 4) the training data in the mini-batch.

To rigorously establish the zero-knowledge property of VeriLoRA, we define two executions from the perspective of any PPT adversary $\mathcal{A}$: a real execution with an honest prover and an ideal execution with a simulator. The security objective is to guarantee that the adversary cannot distinguish these two executions.

Formally, VeriLoRA is zero-knowledge if there exists a polynomial-time simulator $\mathcal{S}$ such that, for all PPT adversaries $\mathcal{A}$ and for all witness $w$ (containing pre-trained parameters, LoRA matrices, intermediate computation values, and training data), the following distributions are computationally indistinguishable: $\left| 
\Pr\left[\mathcal{A}(\mathrm{Real}(w, \mathrm{pp}))=1\right]
-\Pr\left[\mathcal{A}(\mathrm{Ideal}(\mathcal{S}, \mathrm{pp}))=1\right]
\right| \le \mathsf{negl}(\lambda)$, where $\mathrm{Real}(w, \mathrm{pp})$ and $\mathrm{Ideal}(\mathcal{S}, \mathrm{pp})$ denote the real and ideal executions, respectively, and $\mathrm{pp}$ denotes public parameters. The definitions of $\mathrm{Real}$ and $\mathrm{Ideal}$ are given below.

\begin{table}[h]
\centering
\begin{tabular}{|p{8cm}|}
\hline
$\mathrm{Real~Execution}~\mathrm{Real}(w, \mathrm{pp})$:
\begin{enumerate}
\item $com \leftarrow VeriLoRA\text{-}Commit(w; pp)$, where $w$ includes the base model’s initial weights and the initial low-rank matrices used by LoRA.
\item $\pi \leftarrow VeriLoRA\text{-}Prove(com; pp)$, using the zkMat, zkElementProd, zkTranspose, zkSwiGLU, and zkSoftmax protocols.
\item \textbf{return} $com, \pi$
\end{enumerate} \\
\hline
$\mathrm{Ideal~Execution}~\mathrm{Ideal}(\mathcal{S}, \mathrm{pp})$:
\begin{enumerate}
\item $com \leftarrow \mathcal{S}_1(1^\lambda; pp)$, where the simulator generates a commitment without access to the witness $w$.
\item $\pi \leftarrow \mathcal{S}_2(com; pp)$, with oracle access to the correctness of the LoRA fine-tuning procedure.
\item \textbf{return} $com, \pi$
\end{enumerate} \\
\hline
\end{tabular}
\label{tab:real-ideal}

\end{table}



In summary, VeriLoRA ensures that the zero-knowledge error is negligible, guaranteeing that the verifier learns nothing beyond the validity of the statement, while preserving the privacy of the prover’s secret witness.

\section{Experimental Evaluations}\label{Experimental}

To validate the practical feasibility and performance of VeriLoRA, we conducted comprehensive experiments on real-world large language models and datasets. This section outlines the experimental setup, implementation details, and evaluation results.

\subsection{Implementation Details}
The implementation of VeriLoRA adopts a hybrid architecture where CUDA components handle core cryptographic operations and zero-knowledge proof generation, while Python scripts provide high-level orchestration and model-specific interfaces. The CUDA modules implement elliptic curve arithmetic and pairing operations on the BLS12-381 curve \cite{boneh2001short}, cryptographic commitments for model parameters, and GPU-accelerated zero-knowledge protocols for LoRA Fine-tuning on LLMs. The Python layer manages model downloading, file I/O utilities, and LLaMA-specific processing workflows. We evaluated VeriLoRA for LoRA fine-tuning across six Transformer-based LLMs of varying scales, including \textbf{LLaMA-3.2} \cite{touvron2023llama} with 3B and 11B parameters, \textbf{LLaMA-2} \cite{touvron2023llama} with 7B and 13B parameters, and \textbf{OPT} \cite{zhang2022optopenpretrainedtransformer} with 6.7B and 13B parameters, covering models up to 13 billion parameters. We trained the LLMs with the public C4 dataset of English-language text \cite{raffel2020exploring}. Our code is publicly available at \url{https://github.com/liaoguofu/VeriLoRA}.

\subsection{Experimental Setup}

The experiments were conducted on a high-performance computing node with the following configuration: \textbf{Memory}: 503.35 GB of RAM; \textbf{CPU}: 192 cores of AMD EPYC 9654 processor (3.7 GHz with 384 MB cache L3); \textbf{GPU}: NVIDIA A100 GPU with 80 GB of memory. 

\subsection{Rescaling Mechanism for Non-Arithmetic Operations}

To manage range constraints and preserve numerical precision, we adopted a rescaling mechanism for non-arithmetic operation inputs, such as SwiGLU activations and the softmax function. When an input value cannot be directly represented within the range of a single fixed-size lookup table, it is rescaled or decomposed into multiple segments. Each segment requires a separate lookup argument. 

\begin{itemize}
    \item \textbf{Lookup Table Size}: Each table is configured with a size of $2^{16}$, balancing resolution with memory constraints and avoiding excessive table sizes.
    \item \textbf{Softmax Scaling}: For the softmax function within the self-attention mechanism of each Transformer layer, input logits are scaled by a fixed factor $\xi$, resulting in a total scaling factor of approximately $2^{64}$.
    \item \textbf{Segment Decomposition}: Each scaled input is decomposed into $K = 5$ digit segments, with each segment verified by a separate lookup table of size $2^{16}$.
\end{itemize}

\begin{figure*}[t]
  \centering
  \includegraphics[width=1\textwidth]{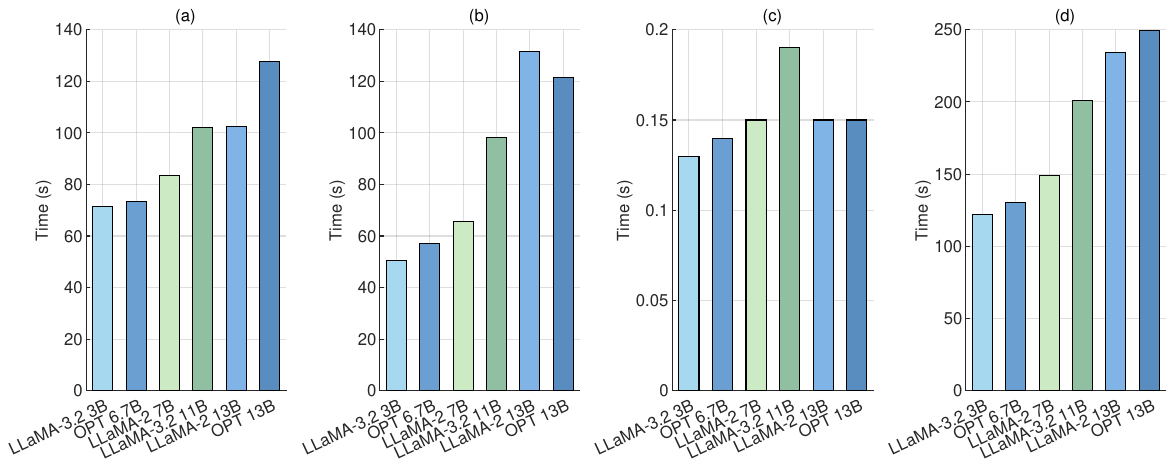}
  \caption{The proving time of VeriLoRA for different models (in seconds): (a) the forward propagation phase; (b) the backward propagation phase; (c) the parameter update phase, (d) the total.}\label{fig 2.png}
\end{figure*}

\subsection{Evaluation Results}

\subsubsection{Proving Time}


We evaluated the prover runtime of VeriLoRA across the six Transformer based LLMs. All measurements correspond to the the training process of a mini-batch consists of a single data sample. For each model, we measured the prover runtime of three distinct computational phases of LoRA: forward propagation, backward propagation, and parameter updates. As shown in Fig.~\ref{fig 2.png}, the total proving time ranges from 121.93 seconds (LLaMA-3.2-3B) to 249.38 seconds (OPT-13B), reflecting the increasing computational overhead as model size scales. In addition, we measure a LoRA fine-tuning mini-batch (without any ZKPs) executes in 0.47 seconds (LLaMA-3.2-1B), 0.89 seconds (LLaMA-3.2-3B), 1.12 seconds (LLaMA-2-7B), and 1.25 seconds (LLaMA-3.1-8B); relative to these no-zk baselines, our VeriLoRA are approximately three orders of magnitude slower. This order-of-magnitude gap reflects the inherent overhead of zero-knowledge proving in practice. For completeness, we also benchmarked a public zkML inference prover: the distilled GPT-2 inference proof generation takes 7,370.6 seconds \cite{chen2024zkml}. Comparing zkML’s inference proof with our inference prover shows a consistent trend—both are orders of magnitude slower than no-ZK execution—while exact ratios vary with model size and prover stack.

\subsubsection{Verification Time}

We measured verification time for the six tested LLMs and present the results in Table~\ref{t2}. Across all tested LLMs, verification remains highly efficient. As shown in Table~\ref{t2}, the end-to-end verification time ranges from 1.87 seconds (LLaMA-3.2-3B) to 3.73 seconds (OPT-13B), consistently staying below 4 seconds even for models with over 13 billion parameters. These results demonstrate that the verification workload in VeriLoRA is much more efficient than the proof generation and represents only a minor fraction of the total computational cost.

\subsubsection{Polynomial Commitment Cost}

We measured the time costs and sizes of the Hyrax polynomial commitments used in VeriLoRA. From the results provided in Table~\ref{t2}, the generation of the commitments introduces a substantial time cost, particularly for large-scale models. The commitment generation time ranges from 156 seconds for LLaMA-3.2-3B to 554 seconds for OPT-13B. This step involves producing polynomial commitments for the different weights of model and dominates the wall-clock latency in our evaluation. The commitment sizes follow a similar trend, increasing from 135.59 MB to 232.67 MB, consistent with the growth in matrix and lookup table dimensionality.

\subsubsection{GPU Memory Usage}

The total memory usage reported in Table~\ref{t2} represents the cumulative GPU memory allocated across all Transformer layers during the forward propagation, backward propagation, and parameter update phases while executing VeriLoRA. These memory usages range from approximately 2573 GB for LLaMA-3.2-3B to 3920 GB for LLaMA-2-11B. The computations for the forward propagation, backward propagation, and parameter update phases within the sub-layers of the Transformer layers are executed sequentially on the GPU. Consequently, the actual peak memory observed during VeriLoRA execution remains below 80 GB. This allows VeriLoRA to be efficiently processed on a single NVIDIA A100 GPU with 80 GB of device memory, ensuring practicality even for the largest model scale evaluated.

\begin{table}[t]
  \centering
  \caption{Verification Time, Polynomial Commitment Cost and GPU Memory Usage for Different Models}\label{t2}
  \renewcommand{\arraystretch}{1.2}
  \begin{adjustbox}{max width=\columnwidth}
  \setlength{\tabcolsep}{6pt}
  \begin{tabular}{l|cccc}
  \toprule
  Model & \makecell{Verification\\Time (s)} & \makecell{Commitment\\Time (s)} &
  \makecell{Commitment\\Size (MB)} & \makecell{Total Memory\\Usage (GB)} \\
  \midrule
  LLaMA-3.2 3B  & 1.87  & 156   & 135.59 & 2573 \\
  LLaMA-3.2 11B & 2.99  & 299   & 228.31 & 3920 \\
  LLaMA-2 7B    & 2.23  & 232   & 182.84 & 2933 \\
  LLaMA-2 13B   & 3.49  & 304   & 224.56 & 3902 \\
  OPT 6.7B      & 1.95  & 285   & 165.48 & 3135  \\
  OPT 13B       & 3.73  & 554   & 232.67 & 3912 \\
  \bottomrule
  \end{tabular}
  \end{adjustbox}

  \end{table}

\section{Related Work} \label{related work}

In this section, we survey verifiable machine learning from three  perspectives:  i) verifiable inference, which focuses on proving the correctness of a model’s forward propagation under zero knowledge;  ii) verifiable training, which aims to authenticate gradient computation and parameter updates; and iii) verifiable testing, which certifies global performance metrics (e.g.\ accuracy or fairness) without revealing model or data. 

\subsection{Verifiable Inference}

Early works like SafetyNets \cite{ghodsi2017safetynets} used interactive proofs to verify DNN predictions without privacy protection. Subsequent zero-knowledge approaches like vCNN \cite{lee2024vcnn} introduced efficient convolution encodings for zk-SNARKs.

For Transformers, ZKML \cite{chen2024zkml} provided the first end-to-end zero-knowledge proof for GPT-2 inference using Halo2 lookup arguments, though with substantial proving overhead. zkLLM \cite{sun2024zkllm} introduced GPU acceleration, reducing latency and scaling to 13B parameters. In addition, zkGPT \cite{qu2025zkgpt} achieved sub-25 second proofs for GPT-2 (117M) through constraint fusion and circuit optimization, outperforming prior work by 185-279×.


More recently, research has extended to the verifiable inference of parameter-efficient adapters, focusing on proving the correct usage of private LoRA weights during the forward pass. Specifically, ZKLoRA \cite{roy2025zklora} addresses distributed or outsourced inference contexts by introducing a multi-party framework utilizing recursive zero-knowledge arguments (i.e., folding schemes) to verify that private LoRA weights are correctly applied during the forward propagation of LLMs. Similarly, ZK-EdgeLoRA \cite{li2025zk} targets edge-cloud collaborative scenarios by employing a VOLE-based commit-and-prove protocol, which uses random linear combinations to efficiently validate matrix multiplications.

Collectively, these efforts demonstrate that ZKPs can authenticate machine learning inference steps—the model owner (prover) can convince a client (verifier) that a given prediction was computed correctly by a neural network, without revealing inputs or weights. However, all of the above works target the inference procedure of fixed, pre-trained models. In contrast, our work focuses on proving the training (fine-tuning) process itself, which involves a series of parameter updates rather than a single forward-pass computation.

\subsection{Verifiable Training}
Verifying training is more demanding than inference due to gradient computation. Early approaches side-stepped this complexity by verifying only snapshots. VeriML \cite{zhao2021veriml} asks the service provider to commit to intermediate states and later supply ZK proofs for randomly sampled iterations, covering linear and shallow-net models but not deep networks.  Distributed settings such as Drynx \cite{froelicher2020drynx} and zkMLaaS \cite{huang2022zkmlaas} combine MPC and interactive proofs to check aggregated updates, again limited to low-complexity learners.

To support full deep neural network (DNN) training, Sun and Zhang introduced zkDL \cite{sun2024zkdl}, which developed a custom zkReLU protocol to prove forward and backward propagation through ReLU activations. zkDL flattens the training process into a parallel circuit architecture (FAC4DNN), enabling the generation of a single-epoch proof for CNNs with tens of millions of parameters in under a minute.

Building on the zkPoT framework, Abbaszadeh \cite{abbaszadeh2024zero} further optimized per-step GKR-based sumchecks and recursively aggregated the results across gradient steps, yielding concise proofs across multiple training iterations for moderate-sized models.

Beyond arithmetic correctness, Shamsabadi et al.\ \cite{shamsabadi2022confidential} introduced Confidential-PROFITT, which provides zero-knowledge proofs of fair training for decision trees. Their protocol certifies that the trained model satisfies demographic parity constraints, without revealing model parameters or sensitive attributes, and completes in minutes.

Despite these advances, no prior work supports verifiable fine-tuning of large-scale Transformer models. In particular, LoRA \cite{hu2022lora}, a widely adopted technique that adapts only low-rank matrices while freezing the base model, has not been addressed in prior ZKP systems. Our work is the first to construct a zero-knowledge proof system for LoRA fine-tuning, verifying that parameter updates are computed correctly on the specified training data while preserving confidentiality.

\subsection{Verifiable Testing}
Verifiable \emph{testing} seeks to certify global properties—such as accuracy or fairness—of a trained model on an evaluation set, while hiding the model and inputs.  The first dedicated ZKP for this task is zkDT \cite{zhang2020zero}, which proves both per-sample inference and dataset-wide accuracy for decision-tree models by embedding all test paths in a single Aurora proof; a 23-level tree evaluated on 5000 CIFAR samples yields a 250s proof of only 287 KB.  Campanelli \emph{et al.} extend this line with cq+/zkcq+ matrix-lookup arguments \cite{campanelli2024lookup}, letting the prover commit to the entire decision tree as a matrix and isolate only the rows reached by all test inputs.  Their zero-knowledge matrix lookup reduces prover time to be independent of the tree size, substantially cutting the overhead of zkDT.

For convolutional networks, pvCNN \cite{weng2023pvcnn} combines homomorphic encryption, Quadratic-Matrix-Program (QMP) circuits, and proof aggregation to batch-verify accuracy across many testers; its QMP encoding yields a 13.9× speed-up over traditional QAP-based zk-SNARKs for high-dimensional convolutions.  ZEN \cite{feng2021zen} introduces a compiler that quantizes floating-point CNNs into R1CS with sign-bit grouping and SIM(D) stranded encoding, reducing constraints by up to 22× with no accuracy loss.  zkCNN \cite{liu2021zkcnn} further optimises two-dimensional convolution and FFT inside Groth16, achieving end-to-end proofs for VGG-16 in 88 s with 341 KB proofs and 59 ms verification.

Collectively, these systems demonstrate that ZKPs can certify model-wide statistics or properties without exposing model weights or evaluation data. However, they all presume access to a fixed, pre-trained model and do not address the fundamental challenge of verifying how the model was trained or fine-tuned. This limitation is critical in scenarios where the training process itself must be auditable and trustworthy. Our work addresses this gap by providing the first zero-knowledge proof system for LoRA fine-tuning of large Transformers, ensuring that the adaptation process is verifiably correct while maintaining privacy.

\section{Conclusion}\label{conclusion}
This work introduces \textbf{VeriLoRA}, the first framework to achieve zero-knowledge verifiability for the fine-tuning of large language models using parameter-efficient methods like LoRA. \textbf{VeriLoRA} provides end-to-end proofs of correctness for the entire fine-tuning process, including forward and backward propagation as well as parameter updates, without disclosing model parameters or training data. By developing new techniques for handling non-arithmetic operations in Transformer architectures and optimizing the proof system for scalability, \textbf{VeriLoRA} enables secure, privacy-preserving model adaptation on billion-parameter models. Experimental results on open-source LLMs demonstrate the practicality and efficiency of the approach, paving the way for trustworthy deployment of fine-tuned LLMs in sensitive or untrusted settings.

Looking ahead, one promising direction is the development of lighter-weight proof systems that further reduce the hardware requirements and prover time. We hope \textbf{VeriLoRA} serves as a foundation for future research at the intersection of cryptography and large-scale machine learning, fostering trustworthy deployment of LLMs in security-critical and privacy-sensitive environments.

\section*{Acknowledgement}
This work was supported in part by the National Natural Science Foundation of China (NSFC) under Grant 62471316, in part by the Shenzhen Key Research Project under Grant JCYJ20220818100810023, and in part by the Program of Science and Technology Cooperation of Nanjing with International/Hong Kong, Macao and Taiwan under Grant 202401019, and in part by the Shenzhen Science and Technology Program under Grant JCYJ20220531101015033.

\bibliographystyle{IEEEtran}
 \bibliography{main} 



\end{document}